# Air-stable, earth-abundant molten chlorides and corrosion-resistant containment for chemically-robust, high-temperature thermal energy storage for concentrated solar power

*Adam S. Caldwell[a], Grigorios Itskos[a] and Kenneth H. Sandhage*[a]*

[a]Purdue University, School of Materials Engineering, West Lafayette, IN 47907, USA

*Corresponding Author

*E-mail addresses:* Sandhage, K.H. (sandhage@purdue.edu), Itskos, G. (gitskos@purdue.edu), Caldwell, A. (caldwe23@purdue.edu)

**A dramatic reduction in man-made $CO_2$ emissions could be achieved if the cost of electricity generated from concentrated solar power (CSP) plants could become competitive with fossil-fuel-derived electricity. The solar heat-to-electricity conversion efficiency of CSP plants may be significantly increased (and the associated electricity cost decreased) by operating CSP turbines with inlet temperatures $\geq750^oC$ instead of $\leq550^oC$, and by using thermal energy storage (TES) at $\geq750^oC$ to allow for rapidly dispatchable and/or continuous electricity production. Unfortunately, earth-abundant $MgCl_2$-KCl-based liquids currently being considered as low-cost media for large-scale, high-temperature TES are susceptible to oxidation in air, with associated undesired changes in liquid composition and enhanced corrosion of metal alloys in pipes and tanks containing such liquids. In this paper, alternative high-temperature, earth-abundant molten chlorides that are stable in air are identified via thermodynamic calculations. The oxidation resistance, and corrosion-resistant containment, of such molten chlorides at $750^oC$ are then demonstrated. Such chemically-robust, low-cost TES media and effective containment provide critical advances towards the higher-temperature operation of, and lower-cost electricity generation from, CSP plants.**

**Introduction**

Concentrated solar power (CSP) technology with thermal energy storage (TES) has the potential to facilitate large-scale penetration of renewable solar energy into the electricity grid [1-3]. Indeed, one of the main competitive attributes of CSP plants is that heat from sunlight can be readily transferred to, and retained in, thermal energy storage (TES) media, such as molten salts, to allow for rapidly dispatchable power and/or continuous power [4]. Molten nitrate salts, such as a non-eutectic salt mixture comprised of 60 wt% sodium nitrate and 40 wt% potassium nitrate (known as "solar salt"), have commonly been used for heat transfer and TES media in CSP installations [5]. However, such nitrate salts are limited in use to temperatures <600°C owing to thermal decomposition into nitrites [5,6]. The drive to raise the inlet temperatures of CSP turbines, for enhanced thermal-to-electrical conversion efficiency [4], has led to consideration of relatively low-cost, earth-abundant molten chloride salts (e.g., $MgCl_2$-KCl-based compositions) as candidates for heat transfer fluids and TES media for next generation CSP plants [7-9]. Unfortunately, such $MgCl_2$-bearing liquids are prone to oxidation and contamination with oxygen-bearing species upon high-temperature exposure to air, with associated undesired changes in the composition and properties of the liquid, as well as in enhanced corrosion of metal alloys used to contain such liquids [10,11]. While tight sealing of pipes and storage tanks, with active salt monitoring and with continuous gettering of oxygen-bearing species (notably with the use of magnesium [12-14]), may be used to reduce such salt contamination and oxidation, these steps would increase the design complexity, and add to the operational cost, of CSP plants. The purpose of this paper is to demonstrate a more cost-effective and reliable strategy for high-temperature TES. With this new strategy, earth-abundant, *air-stable* (oxidation-resistant) molten chlorides are paired with containment materials that are resistant to oxidation *in air* and to dissolution in such molten chlorides in air [15,16]. Such containment materials include metals or alloys possessing protective external oxide scales that grow and dissolve slowly in the molten chlorides in air, and ceramic liners comprised of such oxides. To demonstrate this new strategy, the oxidation resistance of one such air-tolerant molten chloride (a CaCl$_2$-NaCl liquid) at 750°C, and the associated corrosion resistance of one such containment material (Ni with a slow-growing NiO scale), are evaluated in this paper.

**Thermodynamic Calculations**

The following reactions may be considered to evaluate the relative oxidation resistances of MgCl$_2$, CaCl$_2$, SrCl$_2$, BaCl$_2$, NaCl, and KCl in air at 750°C:

$$\text{MgCl}_2 + \tfrac{1}{2}\text{O}_2(g) \leftrightarrow \text{MgO} + \text{Cl}_2(g) \qquad (1)$$
$$\Delta G°_{rxn(1)}(750°C) = -44.3 \text{ kJ mol}^{-1}$$

$$\text{CaCl}_2 + \tfrac{1}{2}\text{O}_2(g) \leftrightarrow \text{CaO} + \text{Cl}_2(g) \qquad (2)$$
$$\Delta G°_{rxn(2)}(750°C) = +143.1 \text{ kJ mol}^{-1}$$

$$\text{SrCl}_2 + \tfrac{1}{2}\text{O}_2(g) \leftrightarrow \text{SrO} + \text{Cl}_2(g) \qquad (3)$$
$$\Delta G°_{rxn(3)}(750°C) = +180.3 \text{ kJ mol}^{-1}$$

$$\text{BaCl}_2 + \tfrac{1}{2}\text{O}_2(g) \leftrightarrow \text{BaO} + \text{Cl}_2(g) \qquad (4)$$
$$\Delta G°_{rxn(4)}(750°C) = +240.2 \text{ kJ mol}^{-1}$$

$$2\text{NaCl} + \tfrac{1}{2}\text{O}_2(g) \leftrightarrow \text{Na}_2\text{O} + \text{Cl}_2(g) \qquad (5)$$
$$\Delta G°_{rxn(4)}(750°C) = +356.0 \text{ kJ mol}^{-1}$$

$$2\text{KCl} + \tfrac{1}{2}\text{O}_2(g) \leftrightarrow \text{K}_2\text{O} + \text{Cl}_2(g) \qquad (6)$$
$$\Delta G°_{rxn(5)}(750°C) = +457.6 \text{ kJ mol}^{-1}$$

The standard Gibbs free energy change for each of these reactions [17,18] at 750°C, $\Delta G°_{rxn}(750°C)$, are also shown above. These $\Delta G°_{rxn}$ values indicate that the oxidation of CaCl$_2$, SrCl$_2$, BaCl$_2$, NaCl, and KCl can be much less favored than for MgCl$_2$ oxidation. Consider, as comparative examples, the oxidation of MgCl$_2$ dissolved in a eutectic MgCl$_2$-KCl melt (32±2 mol% $MgCl_2$; $T_{eut}$ = 433±9°C [8,19-22]) and the oxidation of $CaCl_2$ dissolved in a eutectic $CaCl_2$-NaCl melt (52±2 mol% $CaCl_2$; $T_{eut}$ = 500±8°C [23-27]). Using activity data [28] reported for $MgCl_2$ in a $MgCl_2$(32 mol%)-KCl melt, and assuming a unit activity for MgO(s) and that chlorine and oxygen behave as ideal gases, the equilibrium partial pressure ratio, $p[Cl_2]/(p[O_2])^{1/2}$, for the oxidation of $MgCl_2$ in this melt at 750°C was found to be 3.0. Consequently, $MgCl_2$ oxidation will be favored for this chloride liquid in air (oxygen partial pressure, $p[O_2]$ = 0.21 atm) at 750°C unless the effective equilibrium chlorine partial pressure, $p[Cl_2]$, exceeds 1.4 atm; that is, the $MgCl_2$ dissolved in this $MgCl_2$-KCl liquid should undergo spontaneous oxidation at ambient pressure in air at 750°C. Using activity data for $CaCl_2$ in a $CaCl_2$(53 mol%)-NaCl melt [28], with a unit activity for CaO(s) and assuming ideal gas behavior for chlorine and oxygen, the equilibrium partial pressure ratio, $p[Cl_2]/(p[O_2])^{1/2}$, for the oxidation of $CaCl_2$ in this melt at 750°C was found to be $2.5 \times 10^{-8}$. Consequently, $CaCl_2$ oxidation will not be thermodynamically favored for this chloride liquid in air once the local effective chlorine partial pressure exceeds a value of $1.2 \times 10^{-8}$ atm (12 ppb). Similar calculations using thermodynamic data [17,28] for eutectic $SrCl_2$-NaCl (52 mol% $SrCl_2$, 48 mol% NaCl; $T_{eut}$ = 565°C [29]) and $BaCl_2$-NaCl liquids (40 mol% $BaCl_2$, 60 mol% NaCl; $T_{eut}$ = 651°C [30]) yield even smaller values for the equilibrium chlorine partial pressures required to avoid $SrCl_2$ and $BaCl_2$ oxidation in such liquids, respectively, in air at 750°C. Hence, the oxidation of $CaCl_2$, $SrCl_2$, and $BaCl_2$ present in eutectic $CaCl_2$-NaCl, $SrCl_2$-NaCl, and $BaCl_2$-NaCl liquids, respectively, in air at 750°C are much less strongly favored than the oxidation of $MgCl_2$ dissolved in the eutectic $MgCl_2$-KCl liquid under similar conditions.

In order to utilize such air-stable molten chlorides for high-temperature heat transfer and thermal energy storage for CSP (or other electricity-generating) plants, corrosion-resistant pipes and tanks are required for the containment of such molten salts. Consider the following reactions of NiO, $Cr_2O_3$, and $Al_2O_3$ with chlorine:

$$NiO + Cl_2(g) \leftrightarrow NiCl_2 + \tfrac{1}{2}O_2(g) \quad (7)$$
$$\Delta G°_{rxn(7)}(750°C) = -5.7 \text{ kJ mol}^{-1}$$

$$1/3 Cr_2O_3 + Cl_2(g) \leftrightarrow 2/3 CrCl_3 + \tfrac{1}{2}O_2(g) \quad (8)$$
$$\Delta G°_{rxn(8)}(750°C) = +72.3 \text{ kJ mol}^{-1}$$

$$1/3 Al_2O_3 + Cl_2(g) \leftrightarrow 2/3 AlCl_3 + \tfrac{1}{2}O_2(g) \quad (9)$$
$$\Delta G°_{rxn(9)}(750°C) = +115.5 \text{ kJ mol}^{-1}$$

The calculated chlorine partial pressure associated with the equilibration of $CaCl_2$ present in a molten $CaCl_2$(53 mol%)-NaCl salt with air at 750°C was $1.2 \times 10^{-8}$ atm [17,18,28]. At this equilibrium $p[Cl_2]$ value in air, pure solid NiO, $Cr_2O_3$, and $Al_2O_3$ can undergo reaction with this $CaCl_2$-NaCl liquid only if the activities of the chloride products dissolved in the melt are sufficiently low. The critical activities of $NiCl_2$, $CrCl_3$, and $AlCl_3$ required for such reaction with this $CaCl_2$-NaCl salt in air at 750°C were found to be $7.6 \times 10^{-9}$, $1.1 \times 10^{-17}$, and $6.1 \times 10^{-21}$, respectively [17]; that is, after very small amounts of chloride product formation and dissolution, these oxides should become thermodynamically stable with the eutectic $CaCl_2$-NaCl melt in air at 750°C. Consequently, external NiO, $Cr_2O_3$, or $Al_2O_3$ scales formed upon oxidation of solid metal or alloy pipes and tanks, or such oxides present in ceramic-lined pipes and tanks, should become resistant to reaction with air-saturated $CaCl_2$-NaCl melts after minimal interaction with such liquids.

**Experimental Results and Discussion**

In light of the thermodynamic analyses above, the relative resistances of $MgCl_2$, dissolved in a $MgCl_2$(32 mol%)-KCl liquid, and of $CaCl_2$, dissolved in a $CaCl_2$(53 mol%)-NaCl liquid, towards high-temperature oxidation have been examined by exposing these molten salts to flowing pre-dried air at 750°C (Experimental Methods). Room-temperature X-ray diffraction (XRD) patterns obtained from these salts before and after such air exposure are provided in **Fig. 1**. After only 2.5 h of exposure of the $MgCl_2$-KCl salt to air at 750°C, the presence of appreciable MgO was detected within this solidified salt (**Fig. 1b**) along with KCl and a small amount of the $K_2MgCl_4$ phase (this latter compound was the predominant phase detected in the solidified MgCl$_2$-KCl salt prepared prior to such high-temperature air exposure, **Fig. 1a**) [31]. On the other hand, only diffraction peaks for CaCl$_2$ and NaCl (i.e., not for CaO) were detected in the solidified CaCl$_2$-NaCl salt before and after exposure for 50 h to air at 750°C (**Figs. 1c** and **d**) [31]. Weight change measurements were also used to evaluate the oxidation resistance of the molten CaCl$_2$-NaCl salt. Exposure of the molten CaCl$_2$-NaCl salt to flowing pre-dried Ar or pre-dried air for 12 h at 750°C resulted in similar modest relative weight losses (7±3% in Ar and 4±5% in air). Significant oxidation of the CaCl$_2$ or NaCl (i.e., the replacement of heavier chlorine with lighter oxygen) in this molten salt would have resulted in greater weight losses in air than in Ar (with some weight loss in both cases also attributed to salt evaporation). The lack of detectable oxidation of the CaCl$_2$-NaCl molten salt (from XRD analysis and weight change measurements), and the extensive oxidation of MgCl$_2$ present in the MgCl$_2$-KCl melt, were consistent with the thermodynamic calculations above; that is, the molten CaCl$_2$-NaCl salt was significantly more resistant to oxidation in air at 750°C than the molten MgCl$_2$-KCl salt.

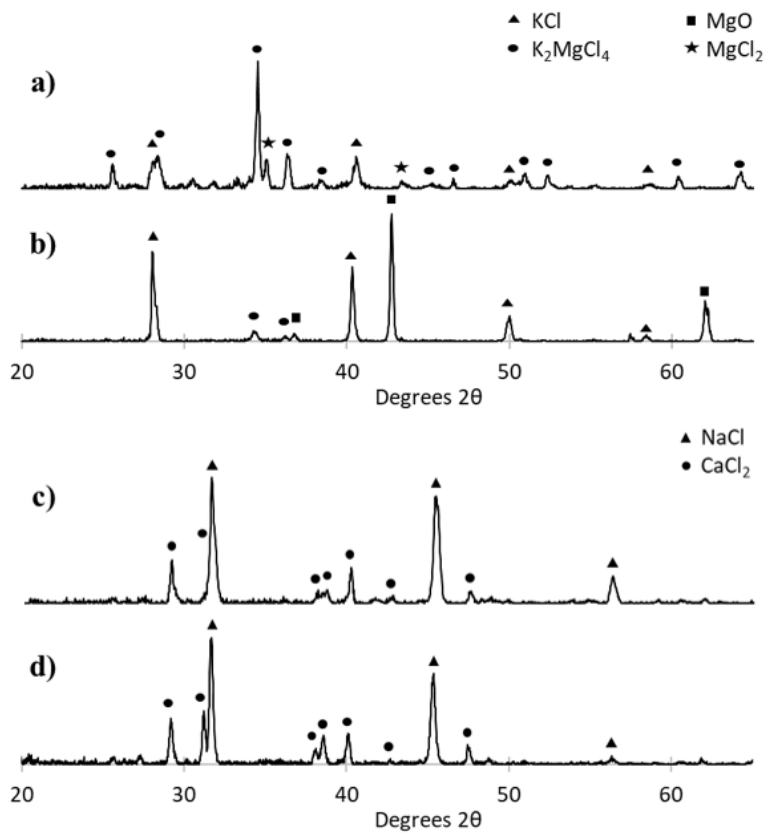

**Figure 1.** Relative stabilities of molten $MgCl_2$-KCl and $CaCl_2$-NaCl salts in air at 750°C. Room temperature X-ray diffraction patterns obtained from (a), (b) a $MgCl_2$(32 mol%)-KCl salt after exposure to pre-dried, high-purity Ar for 50 h at 750°C or to pre-dried air for 2.5 h at 750°C, respectively, and (c), (d) a $CaCl_2$(53 mol%)-NaCl salt after exposure to pre-dried, high-purity Ar for 50 h at 750°C or to pre-dried air for 50 h at 750°C, respectively

To demonstrate the corrosion-resistant containment of the air-stable molten CaCl$_2$-NaCl salt, pre-oxidized, plate-shaped Ni specimens (with continuous external NiO scales) were exposed (in a vertical orientation) to molten CaCl$_2$-NaCl salts in ambient air at 750°C for up to 48 h. Two types of molten salts were utilized in these corrosion experiments, in order to examine the influence of the starting nickel content of the salt on such corrosion: CaCl$_2$ (53 mol%)-NaCl salts that had been presaturated with NiO at 750°C in air, and CaCl$_2$(53 mol%)-NaCl without such NiO presaturation. Inductively-coupled plasma-mass spectroscopic analyses indicated that the NiO-saturated molten CaCl$_2$-NaCl salt contained a nickel content of 27.5±9.7 ppm (atomic basis, **Fig. S1**, ESI). Plots of the NiO/Ni specimen mass change per area, $\Delta m\ A^{-1}$, as a function of immersion time in these molten salts are shown in **Figs. 2a** and **b**. The pre-oxidized Ni specimens exhibited a modest mass gain with immersion time in the molten $CaCl_2$-NaCl salt at 750°C in air. Similar $\Delta m\ A^{-1}$ values were obtained for corrosion of these specimens in molten $CaCl_2$-NaCl salts with or without presaturation with NiO. As shown in **Fig. 2b**, a good fit was obtained to parabolic kinetics ($\Delta m\ A^{-1}$ vs. $t^{1/2}$), with the best fit line yielding yielded a slope of $6.0\pm0.3 \times 10^{-7}$ g $cm^{-2}$ $s^{-1/2}$ ($3.6\pm0.2 \times 10^{-2}$ mg $cm^{-2}$ $h^{-1/2}$). This value was not far from the measured rate of mass change per area for NiO-bearing (pre-oxidized) Ni specimens exposed only to air at 750°C. Thermogravimetric analysis of pre-oxidized Ni in flowing synthetic dry air at 750°C yielded a parabolic plot with a fitted slope of $8.2\pm0.9 \times 10^{-7}$ g $cm^{-2}$ $s^{-1/2}$ ($4.9\pm0.5 \times 10^{-2}$ mg $cm^{-2}$ $h^{-1/2}$). The following expression has also been reported for the rate of oxidation of Ni specimens (at $\geq1100$°C) that had first been pre-oxidized at 1400°C [32] (i.e., a similar pre-oxidation temperature as used in the present work):

$$k_p = 0.377(p[O_2])^{1/12} \exp[-119.5\ kJ\ mol^{-1}\ R^{-1}\ T^{-1}]\ g\ cm^{-2}\ s^{-1/2} \qquad (10)$$

The extrapolated $k_p$ value obtained from this equation for air at 750°C (i.e., $p[O_2] = 0.21$ atm, 1023 K) was $2.6 \times 10^{-7}$ g $cm^{-2}$ $s^{-1/2}$. This extrapolated value was also not far (within factor of 2.3) from the corrosion rate obtained in molten $CaCl_2$-NaCl salt in the present work. The similarity in the observed rates for the oxidation of NiO/Ni specimens in air at 750°C and for the corrosion of NiO/Ni specimens in the molten $CaCl_2$-NaCl salt in air at 750°C was consistent with a similar rate-controlling mechanism for the oxidation of NiO-bearing Ni specimens in both environments, namely, solid-state diffusion through the NiO scale (via outward Ni cation migration) [32]. A backscattered electron (BSE) image, and associated elemental maps (from energy-dispersive X-ray (EDX) analyses), of a polished cross-section of the interface between a pre-oxidized Ni specimen and solidified $CaCl_2$-NaCl salt, after exposure of the specimen to the molten salt for 48 h in air at 750°C, is shown in **Fig. 3**. The elemental maps indicated that the molten $CaCl_2$-NaCl salt did not penetrate through the NiO scale, so that continued oxidation of the underlying Ni exposed to this molten salt in air at 750°C required solid-state diffusion through the NiO scale. The parabolic corrosion rate obtained for the pre-oxidized Ni specimens in the molten $CaCl_2$-NaCl salt in air at 750°C corresponded to an annual mass change per area value of $3.4 \times 10^{-3}$ g cm$^{-2}$, or an annual Ni recession (with the molar volume of Ni = 6.59 cm$^3$ mol$^{-1}$ [31]) of only 14 μm.

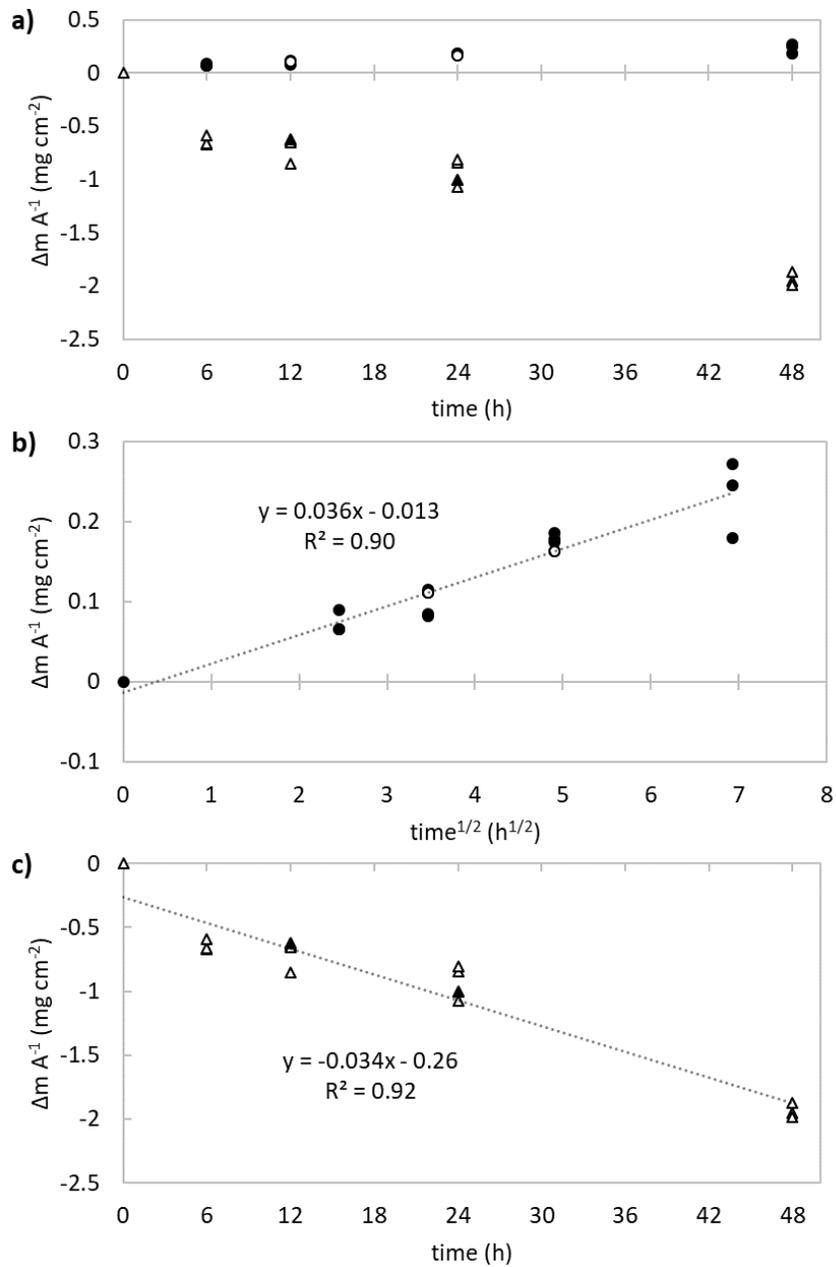

**Figure 2.** Corrosion kinetics of NiO/Ni (pre-oxidized Ni) and Ni specimens in molten $CaCl_2$(53 mol%)-NaCl salts in air at 750°C. (a) $\Delta m/A$ vs. time plot for exposure of NiO/Ni (circles) and as-polished Ni (triangle) specimens to the molten $CaCl_2$-NaCl salt that had (closed symbols) or had not (open symbols) been pre-saturated with NiO. (b) Parabolic ($\Delta m/A$ vs. (time)$^{1/2}$) plot for exposure of NiO/Ni specimens to the molten $CaCl_2$-NaCl salt. (c) Linear ($\Delta m/A$ vs. time) plot for exposure of as-polished Ni specimens to the molten $CaCl_2$-NaCl salt.

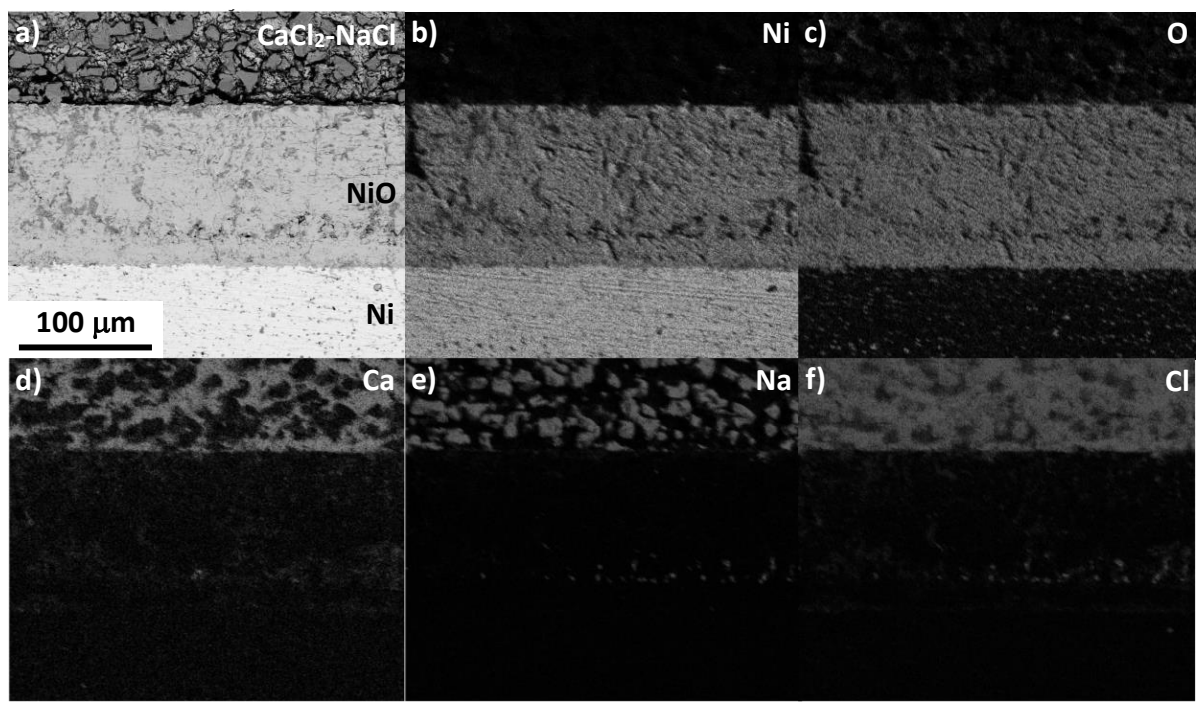

**Figure 3.** The interface between a pre-oxidized Ni specimen and solidified $CaCl_2$-NaCl liquid. (a) BSE image and (b)-(f) associated elemental maps for Ni, O, Ca, Na, and Cl, respectively, obtained from a polished cross-section of a pre-oxidized Ni (NiO/Ni) sample that had been exposed to a $CaCl_2$(53 mol%)-NaCl liquid for 48 h in ambient air at 750°C.

Experiments were also conducted to evaluate the corrosion of as-polished Ni specimens (i.e., specimens that had not been exposed to a pre-oxidation thermal treatment) in the molten $CaCl_2$-

NaCl salt in air at 750°C. Plots of the as-polished Ni specimen mass change per area, $\Delta m\, A^{-1}$, as a function of immersion time in this molten salt are shown in **Figs. 2a** and **c**. Unlike the pre-oxidized (NiO/Ni) specimens, these Ni specimens exhibited a significant mass loss with immersion time. The mass loss per area with time was found to exhibit a good fit to linear kinetics (**Figs. 2c**), with similar $\Delta m\, A^{-1}$ values obtained for corrosion of these specimens in molten $CaCl_2$-NaCl salts without and with presaturation with NiO. Backscattered electron (BSE) images, and elemental maps (from EDX analyses), of a polished cross-section of the interface between a Ni specimen and solidified $CaCl_2$-NaCl salt, after exposure of the specimen to the molten salt for 48 h in air at 750°C, are shown in **Fig. 4**. A continuous NiO layer was not detected at the interface between the

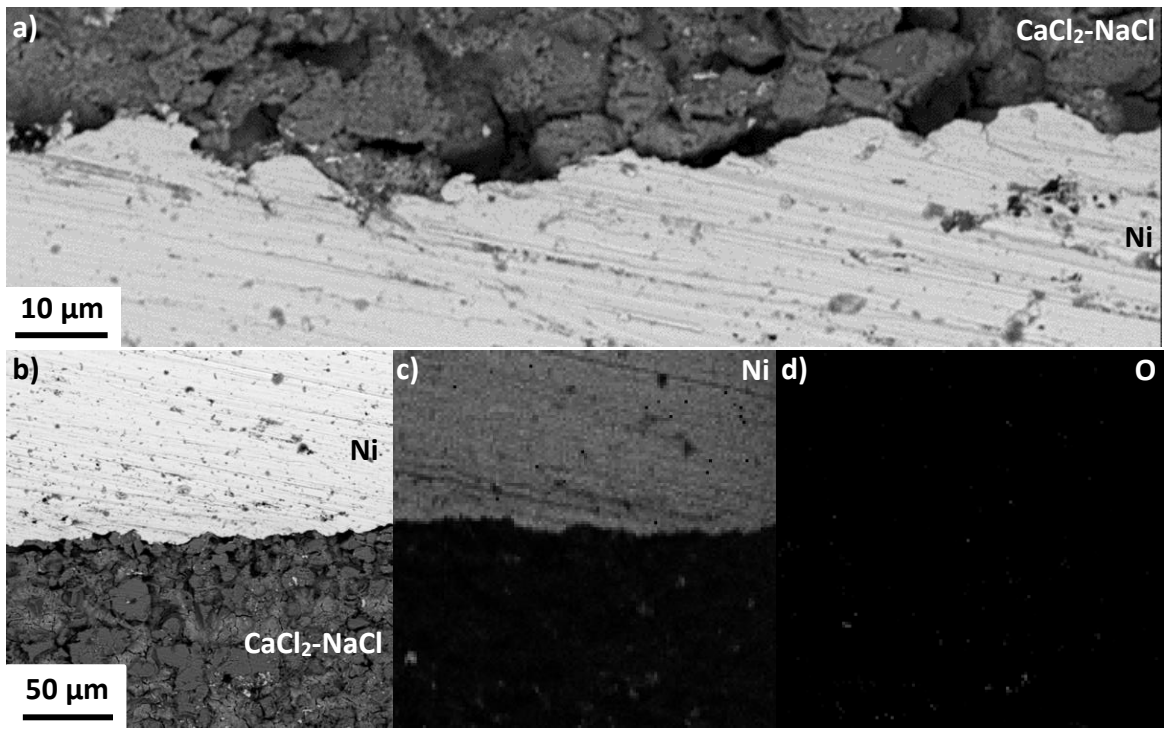

**Figure 4.** The interface between a Ni specimen and solidified $CaCl_2$-NaCl liquid. (a), (b) BSE images and (c), (d) associated elemental maps for Ni and O, respectively, obtained from a polished cross-section of an as-polished Ni sample that had been exposed to a $CaCl_2$(53 mol%)-NaCl liquid for 48 h in ambient air at 750°C.

Ni specimen and solidified CaCl$_2$-NaCl salt. Although the formation of NiO was thermodynamically favored (as discussed above), the nucleation and growth of a continuous external NiO scale on the Ni surface was apparently kinetically inhibited in the presence of the molten CaCl$_2$-NaCl salt. The resulting direct exposure of the Ni specimens to the molten CaCl$_2$-NaCl salt, and the observed specimen mass loss with time, were consistent with the corrosion of Ni via reaction and dissolution into the molten CaCl$_2$-NaCl salt. Indeed, relatively rapid rates of such Ni reaction and dissolution into the molten salt (**Figs. 2a** and **c**), compared to the rates of NiO nucleation and sideways growth at the Ni/molten salt interface, would inhibit the formation of a continuous NiO scale on the Ni surface.

This work demonstrates that air-stable, CaCl$_2$-NaCl-based liquids may be contained (e.g., in pipes and tanks) within corrosion-resistant materials that possess continuous, external oxide scales that exhibit slow (parabolic) growth kinetics in the salt in air at 750°C (or within containment materials comprised entirely of such oxides). The thermodynamic calculations above indicated that this corrosion-resistant containment concept for air-stable molten chlorides may be extended, beyond NiO-bearing Ni, to metal alloys capable of forming other slow-growing external oxide scales in air at 750°C, such as Cr$_2$O$_3$-bearing or Al$_2$O$_3$-bearing alloys (e.g., oxidation-resistant Fe-based or Ni-based alloys) [15,16]. The thermodynamic calculations above also indicated that other multicomponent, air-stable molten chloride salts (beyond just the CaCl$_2$-NaCl salt of the present work) may be utilized for reliable (air-tolerant) heat transfer and TES (i.e., ternary to quintenary salts comprised of CaCl$_2$, SrCl$_2$, BaCl$_2$, NaCl, and/or KCl). For example, lower-melting CaCl$_2$-NaCl-BaCl$_2$ and CaCl$_2$-NaCl-BaCl$_2$-KCl salts (with reported liquidus temperatures of 440-454°C [27,30,33,34] and 421°C [35] respectively) may be utilized. It is also worth noting that such earth-abundant chloride salts are less expensive than solar salt currently used at ≤550°C [10].

**Concluding Remarks**

In summary, this work provides alternative, oxidation-resistant, high-temperature (750°C) fluids for reliable (air-compatible) heat transfer and thermal energy storage (TES), and a corrosion-mitigation strategy for high-temperature containment of such fluids, to allow for higher-temperature operation of, and lower-cost dispatchable or continuous electricity production from, Concentrated Solar Power (or other electricity-producing) plants [15,16]. Thermodynamic calculations indicated that, unlike $MgCl_2$-bearing molten salts currently under consideration for high-temperature heat transfer and TES, molten salts comprised of $CaCl_2$, $SrCl_2$, $BaCl_2$, NaCl, and/or KCl are much more resistant to oxidation in air at 750°C. Indeed, while appreciable oxidation of the $MgCl_2$ in a binary eutectic $MgCl_2$(32 mol%)-KCl melt (yielding solid MgO) occurred within only 2.5 h of exposure to air at 750°C, no oxide products were detected for a binary eutectic $CaCl_2$(53 mol%)-NaCl melt exposed for 50 h to air at this temperature. The choice of robust, oxidation-resistant molten salt compositions for reliable heat transfer or TES may be extended beyond the binary $CaCl_2$-NaCl molten salt demonstrated in this work to include a wide variety of ternary to quintenary molten chlorides comprised of $CaCl_2$, $SrCl_2$, $BaCl_2$, NaCl, and/or KCl that can be selected for tailored properties (such as reduced liquidus temperatures, relative to the eutectic $CaCl_2$-NaCl liquid). Corrosion-resistant containment (e.g., for pipes and tanks) of such air-stable molten chlorides may, in turn, be achieved with the use of metals and alloys capable of forming slow-growing external oxides scales in air at 750°C (or with pipes and tanks lined with such oxides) [15,16]. This has been demonstrated by immersing NiO-bearing Ni specimens immersed in the eutectic $CaCl_2$(53 mol%)-NaCl melt in air at 750°C. These specimens exhibited a slow parabolic rate of mass gain with time that was consistent with solid-state, diffusion-controlled thickening of the NiO scale (at a rate corresponding to an annual Ni recession of 14 µm). Thermodynamic calculations indicate that this concept for containment of air-stable molten chlorides may be extended beyond NiO-bearing Ni to metal alloys capable of forming other slow-growing external oxide scales in air at 750°C, such as Fe-based or Ni-based alloys capable of forming continuous, external $Cr_2O_3$-based or $Al_2O_3$-based scales.

**Experimental Methods**

***Preparation of the $MgCl_2$-KCl and $CaCl_2$-NaCl salt mixtures.*** Chloride salts (NaCl, KCl, $MgCl_2$, and $CaCl_2$, 99% anhydrous, ≥50 mesh particle size, Sigma-Aldrich Corp., St. Louis, MO USA) were individually pre-dried in a vacuum oven (DZF-6050 oven, Hardware Factory Store, Covina, AZ USA) at 30 mTorr for 12 h at 200°C. The chloride salts were then mixed to obtain binary chloride compositions of $CaCl_2$(53.0±0.9 mol%)-NaCl(48.0±0.9 mol%) and $MgCl_2$(32.0±0.9 mol%)-KCl(68.0±0.9 mol%) and stored in an Ar-atmosphere glovebox, with oxygen and water contents in the glovebox maintained at ≤1 ppm and ≤0.5 ppm, respectively.

***Molten salt oxidation tests.*** Pre-dried mixtures of 6 g of the $CaCl_2$(53 mol%)-NaCl composition and 6 g of the $MgCl_2$(32 mol%)-KCl composition were placed in alumina crucibles (1.6 cm internal diameter x 2.4 cm internal height, ≥99.6% purity, Advalue Technology, Tucson, AZ, USA). After sealing the salt-bearing crucibles in a controlled-atmosphere horizontal mullite tube (7.0 cm internal diameter) furnace (STT-1500C-3-36, Sentro-Tech, Strongsville, OH USA), the salts were heated at 3°C/min to 750°C in flowing (1 L/min) Ar (99.999% purity, Airgas Co., Danville, IL USA) or flowing synthetic air (21% $O_2$/79% $N_2$, Airgas Co.). Prior to introduction into the furnace, each gas was passed through a column of Ascarite II (5 cm long by 6.6 cm diameter) (Sigma-Aldrich Corp.) and Drierite (23.8 cm long by 6.6 cm diameter) (Sigma-Aldrich Corp.) to remove residual $CO_2$ and $H_2O$, respectively. For the thermal treatments conducted in flowing Ar, both types of salts were held at 750°C for 50 h, and then cooled at 3°C/min to room temperature. For the thermal treatments conducted in flowing synthetic air, the $MgCl_2$-KCl salt was held for only 2.5 h at 750°C (owing to extensive oxidation within even this modest time), whereas the $CaCl_2$-NaCl salt was held for 50 h at 750°C, prior to cooling at 3°C/min to room temperature.

Weight change measurements of molten $CaCl_2$-NaCl salts in Ar and air atmospheres were also conducted. Alumina crucibles were pre-heated to 800°C at 3°C/min in flowing (1 L/min) Ar (99.999% purity, Airgas). Pre-dried mixtures of 3 g of the $CaCl_2$(53 mol%)-NaCl composition were added to the crucibles positioned in a controlled-atmosphere horizontal mullite tube (7.0 cm internal diameter) furnace (STT-1500C-3-36, Sentro-Tech). The salt-bearing crucibles were then heated at 3°C/min to 750°C in flowing (1 L/min) pre-dried Ar (99.999% purity, Airgas Co.) or in flowing pre-dried synthetic air (21% $O_2$/79% $N_2$, Airgas Co.) and held isothermally for 12 h. Weight change measurements were obtained from the salt-bearing crucibles after such Ar or air exposures. A total of 9 oxidization experiments were completed for each gas (i.e., three replications of two pairs for each gas type).

*Preparation of Ni and Ni/NiO plates*. Nickel (≥99.0% Ni, ASTM B162-99, McMasterCarr, Aurora, OH USA) specimens were cut, via wire electrodischarge machining (FX20K, Mitsubishi Electric Corp., Ratingen, Germany), into plates (14 mm x 10 mm x 1.6 mm) and then polished to a 5 µm finish. Polishing was completed using SiC pads of progressively finer grit size, from 320 grit to 600 grit to 1200 grit (Silicon Carbide Paper, Allied High Tech, Rancho Dominguez, CA, USA), using water for lubrication. A given sample was polished in a single direction and rotated 90 degrees between each change of SiC pad grit size, with polishing at a given SiC pad grit size conducted until the polishing marks of the previous stage were removed. The specimens were then cleaned via ultrasonication for 10 min in ethanol (HPLC grade, 99.8% min purity, Fisher Scientific), followed by rinsing with acetone (HPLC grade, ≥99.5%, Fisher Scientific). To generate a continuous oxide scale on some of the Ni specimen surfaces (the "preoxidized Ni specimens") prior to exposure to the molten $CaCl_2$-NaCl salt, the polished/cleaned Ni specimens were heated in flowing synthetic air (1 L/min, 21% $O_2$/79% $N_2$) at 3°C/min to 1400°C and held at this temperature for 90 min, followed by cooling at 3°C/min. During this oxidation treatment, the polished Ni plates were placed in a tilted orientation in an alumina tray (2 cm width x 10 cm length x 1.5 cm height, ≥99.6% purity, AdValue Technology, Tucson, AZ USA) to allow for ready access of air to all of the plate surfaces.

*Tests of NiO saturation in molten $CaCl_2$-NaCl salt.* The NiO solubility in molten $CaCl_2$-NaCl salts at 750°C in air was determined via inductively-coupled plasma-mass spectroscopic (ICP-MS) analyses. For each saturation test, excess NiO powder (0.1 g, ≥99% purity, <325 mesh, Alfa Aesar, Haverhill, MA USA) was placed at the bottom of a magnesia crucible (99.2 wt% MgO, 0.8 wt% $Y_2O_3$, 1.6 cm internal diameter 2.4 cm internal height,, Tateho Ozark Technical Ceramics, Webb City, MO USA). A pre-dried $CaCl_2$ (53 mol%)-NaCl salt mixture (6 g) was then placed on top of the NiO powder within the magnesia crucible. The NiO/$CaCl_2$-NaCl-bearing crucible was heated at 15°C/min to 750°C in ambient air in a box furnace (SC-2, Paragon Industries, Mesquite, TX USA) and held at this temperature for a time of 12 h, 24 h, or 48 h. The salt was then quenched by removal of the crucible from the hot furnace. After solidification, the bottom 1 cm length of the crucible (where the excess NiO had been placed) was cut away by using with a low-speed saw (200 rpm, TechCut 4™, Allied High Tech, ) with a diamond metal-bonded wafering blade (Allied High Tech) using kerosene (reagent grade, Sigma-Aldrich Corp.) as the cutting fluid (i.e., to avoid dissolving the salt during the cutting process). The remaining salt was mechanically removed (chiseled, scraped) from the crucible and was dissolved in 15.8 M nitric acid (TraceMetal Grade, < 1 ppb metal contamination, Fisher Scientific). The nickel content of the solution was then evaluated by ICP-MS analysis (X Series II Thermo Electron Corp., Waltham, MA USA) at the Center for Applied Isotope Studies at the University of Georgia. A total of 9 ICP-MS analyses were conducted (i.e., 3 repetitions at each 750°C annealing time of 12 h, 24 h, and 48 h).

*Ni and Ni/NiO corrosion tests in molten $CaCl_2$-NaCl and molten NiO-saturated $CaCl_2$-NaCl.* Prior to conducting corrosion tests of the polished Ni and the pre-oxidized Ni specimens at 750°C in air, the $CaCl_2$(53 mol%)-NaCl mixtures were thermally treated in ambient air at 750°C for 24 h in magnesia crucibles (99.2 wt% MgO, 0.8 wt% $Y_2O_3$, 1.6 cm internal diameter 2.4 cm internal height, Tateho Ozark Technical Ceramics). NiO-saturated $CaCl_2$-NaCl salts were prepared by adding excess NiO powder (0.1 g, ≥99% purity, <325 mesh, Alfa Aesar) to the bottom of the crucible prior to introducing 9 g of the $CaCl_2$-NaCl mixture. The mixture was then heated for 24 h in ambient air at 750°C. Four types of corrosion tests were conducted with: i) polished, unoxidized Ni specimens immersed in molten $CaCl_2$-NaCl salt, ii) polished, unoxidized Ni specimens immersed in molten NiO-saturated $CaCl_2$-NaCl salt, iii) pre-oxidized Ni specimens immersed in molten $CaCl_2$-NaCl salt, and iii) pre-oxidized Ni specimens immersed in molten, NiO-saturated $CaCl_2$-NaCl salt. Such corrosion tests were conducted in ambient air at 750°C for varied times. For each test, a plate-shaped specimen (polished Ni or pre-oxidized Ni) was placed vertically in a magnesia crucible (99.2 wt% MgO, 0.8% $Y_2O_3$, 1.6 cm internal diameter, 2.4 cm internal height, Tateho Ozark Technical Ceramics) along with 6 g of the salt ($CaCl_2$-NaCl salt or NiO-saturated $CaCl_2$-NaCl salt) to an approximate depth of the molten salt of 1.5 cm (i.e., each specimen was fully immersed upon melting of the salt). The specimen/salt-bearing crucibles were heated in a box furnace (Paragon SC-2, Strongsville, OH USA) in ambient air at 10°C/min to 750°C, held at this temperature for a time of 6 h to 48 h, and then cooled at 10°C/min to room temperature. Some of the plate-shaped specimens were extracted from the solidified salt via dissolution of the salt in purified de-ionized water at room temperature for 2 h. These specimens were then dried under 30 mTorr vacuum at room temperature for 1 h. The mass of each specimen was then measured using an electronic microbalance (2 µg reproducibility, ME 36S Sartorius Lab Instruments GmbH, Goettingen, Germany), to allow for comparison to the specimen mass measured prior to the corrosion test. The dimensions of each specimen were also measured using a digital micrometer ($\pm 2.5 \times 10^{-4}$ cm accuracy, 3732XFL-2, L. S. Starrett Co, Athol, MA USA), to allow for determination of the mass change per specimen surface area. For other plate-shaped specimens, cross-sections were obtained of the specimen/salt interface by cutting through the MgO crucible containing the frozen salt and sample with a low-speed saw (200 rpm, TechCut 4™, Allied High Tech) with a diamond metal-bonded wafering blade (Allied High Tech) using kerosene (reagent grade, Sigma-Aldrich Corp.) as the cutting fluid (i.e., to avoid dissolving the salt during the cutting process). These cross-sections were then polished to a 6 µm finish in an Ar-atmosphere glovebox with oxygen and water contents maintained at $\leq$1 ppm and $\leq$0.5 ppm, respectively. Such polishing was conducted using a series of diamond-impregnated polishing pads with progressively finer average particle sizes of 30 µm, 15 µm, and 6 µm (Dia-Grid Diamond Metal Plates, Allied High Tech), using hexane (reagent grade, Sigma-Aldrich Corp.) as a lubricant. Manual polishing was conducted for approximately 30 min at each average diamond particle size and rinsed between with excess hexane (prior to using the next average diamond particle size) to remove debris from the previous polishing step. The polished samples were individually placed within a sealed bag (16.5 cm x 14.9 cm Ziploc Sandwich Bag, S. C. Johnson & Son, Racine, WI USA) that was placed inside a second similar sealed bag inside the glovebox. Such double bag sealing was used to keep the specimens in an inert atmosphere for transport to the electron microscope, and removed from the bags just immediately prior to introduction into the vacuum chamber of the microscope.

***Oxidation of Ni/NiO in air.*** *Thermogravimetric analyses (TGA).* Pre-oxidized Ni specimens were heated (STA 449 F1 Jupiter, Netzsch-Gerätebau GmbH, Selb, Germany) in flowing synthetic air (40 mL/min, 21% $O_2$/79% $N_2$) that was passed through a column of Ascarite II (5 cm long by 6.6 cm diameter) (Sigma-Aldrich Corp.) and Drierite (23.8 cm long by 6.6 cm diameter) (Sigma-Aldrich Corp.) at 15°C/min to 750°C and held at this temperature for 24 h, followed by cooling at 15°C/min. The sample was positioned vertically in an alumina crucible (5mL volume, 99.7% purity, Netzsch). Prior to heating, the system was evacuated to 30 mTorr and backfilled with synthetic air three times. A background correction was completed using an identical experimental procedure with an empty alumina crucible.

***Characterization of the salts and metal specimens.***

*X-Ray diffraction (XRD) analyses.* XRD analyses of solidified $MgCl_2$-KCl and $CaCl_2$-NaCl salts, after these salts were annealed at 750ºC in flowing Ar (for 50 h) or in flowing air (for 2.5 h with $MgCl_2$-KCl, for 50 h with $CaCl_2$-NaCl), were conducted using Cu K$\alpha$ radiation (D2 diffractometer, Bruker AXS, Inc., Madison, WI USA) with a scan rate of 4.56º/min.

*Electron microscopic analyses.* Backscattered electron images of the polished cross-sections and energy dispersive X-ray (EDX) analyses (for elemental mapping) were acquired using a field-emission-gun scanning electron microscope (Quanta 3D FEG SEM, FEI Co., Hillsboro, OR USA) operating at an acceleration voltage of 10 keV.

**Notes**

K.H.S. is an inventor on patent applications related to this work that have been filed by (and are owned by) Purdue University. The remaining authors declare no conflict of interest.

**Contributions**

K.H.S. identified the air-stable molten salts and conceived of the corrosion mitigation concept. A.C. developed the experimental design with significant input from G.I. and K.H.S. A.C. conducted the high-temperature experiments and the materials characterization. K.H.S. supervised the research with significant input from G.I. All of the authors contributed to the manuscript writing.

**Supplementary data**

Supplementary data to this article can be found online at:

**Acknowledgments**

This work was supported by the U.S. Department of Energy, Office of Energy Efficiency and Renewable Energy (Award Number DE-EE0008532).

**Supplementary Data**

# Air-stable, earth-abundant molten chlorides and corrosion-resistant containment for chemically-robust, high-temperature thermal energy storage for concentrated solar power

Adam Caldwell, Grigorios Itskos, Kenneth H. Sandhage
Purdue University, School of Materials Engineering, West Lafayette, IN, USA

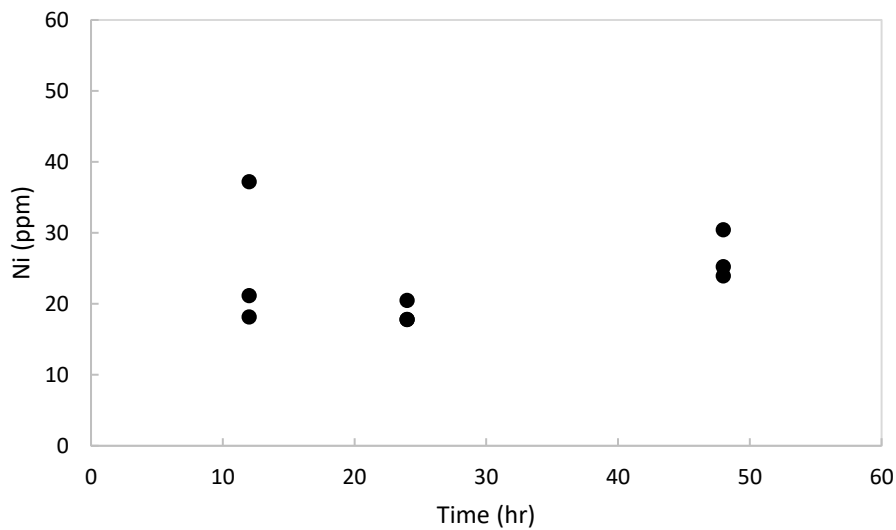

**Figure S1.** Ni concentration (ppm) in the 53 mol% $CaCl_2$/47 mol% NaCl mixtures exposed to excess NiO after thermal treatment for three different times (12 h, 24 h, and 48 h), as determined by ICP-MS analyses.